\documentclass[aps,prb,amsmath,amssymb,reprint,superscriptaddress]{revtex4-2}

\usepackage{graphicx}
\usepackage{amsmath,amssymb}
\usepackage{xcolor}
\usepackage{color,soul}

\begin{document}

\title{Structural and Bonding Insights into B$_8$Cu$_3^-$ Clusters: A DFT Study}

\author{P. L. Rodr\'iguez-Kessler}
\email{plkessler@cio.mx}
\affiliation{Centro de Investigaciones en \'Optica A.C., Loma del Bosque 115, Lomas del Campestre, Leon, 37150, Guanajuato, Mexico}

\date{\today}

\begin{abstract}
In this study, we employ density functional theory (DFT) to investigate the structural and electronic properties of B$_8$Cu$_3^-$ clusters—boron-based frameworks doped with three copper atoms. The results indicate that the lowest-energy structure features a vertical Cu$_3$ triangle supported on a B$_8$ wheel geometry, whereas the horizontally supported configuration is 3.0 (5.6) kcal/mol higher in energy at the PBE0 ($\omega$B97X) functional. Electron localization function (ELF) and Mulliken population analyses reveal that the most stable isomer exhibits strong Cu–B interactions and significant electron delocalization, which contribute to its enhanced stability. Localized orbital locator (LOL) maps further support this finding by showing pronounced electron localization around the Cu$_3$ unit in the more stable structure. These insights highlight the possible role of Cu-centered multicenter bonding in stabilizing boron-based nanoclusters.
\end{abstract}


\maketitle

\section{Introduction}

Boron clusters remain a focal point of research owing to their exceptional structural versatility and unconventional bonding, which set them apart from their carbon-based counterparts. Incorporating transition or rare-earth metals into these clusters has expanded the possibilities for engineering advanced nanomaterials with customized functionalities.\cite{C6CC09570D,C9CP03496J,C9NJ06335H} Notably, metal-doped boron clusters display intriguing electronic and geometric features, including planar aromaticity and metal-centered coordination frameworks.\cite{doi:10.1021/acs.inorgchem.7b02585,doi:10.1021/acs.jpclett.0c02656,doi:10.1021/acs.jpca.1c02148} Of particular interest are inverse sandwich configurations—structures in which two or more metal atoms are positioned between boron rings—due to their pronounced stability and promising implications for materials science applications.

While research on boron clusters doped with two metal atoms (B$_7$M$_2$ systems) has revealed intriguing structural and electronic properties, particularly with inverse sandwich configurations,\cite{D0NJ03999C,D5CP01078K,POZDEEV2023107248,RODRIGUEZKESSLER2025117486,molecules28124721,molecules28124721,B7Y2cluster} studies specifically focused on B$_8$M$_3$ clusters remain limited. In a recent study, a boron-based composite cluster, B$_8$Al$_3^{+}$, doped with three aluminum atoms, was found to have a global minimum structure comprising three layers: an Al$_2$ unit, a B$_8$ ring, and a fluxional isolated Al atom, as revealed by BOMD simulations. This provides a new example of dynamic structural fluxionality in B$_8$M$_3$ systems.\cite{molecules29245961}

Copper–boron systems have attracted significant attention due to their unique electronic and catalytic properties. These systems show promise for applications in nanotechnology, catalysis, and the development of advanced functional materials with tailored electronic and magnetic behaviors. Despite the growing interest in boron-containing compounds, this study focuses on exploring the properties of B$_7$Cu$_3$ clusters—a boron framework doped with three copper atoms—to investigate their structural preferences using density functional theory (DFT).

Although several bonding motifs are conceivable for this system, our comprehensive structural search identifies the Cu$_3$-supported-B$_8$ configuration as the global minimum. To further characterize these structures, we analyze their electron localization functions (ELFs) and localized orbital locator (LOL) maps, offering deeper insights into their bonding nature and overall stability. The information presented in this preprint forms a foundation for future studies on the potential applications of these clusters.\cite{OLALDELOPEZ2025419,https://doi.org/10.1002/adts.202100043}

\section{Computational Details}

All calculations in this study were performed using density functional theory (DFT) as implemented in the ORCA 6.0.0 software package.\cite{10.1063/5.0004608} The exchange–correlation energy was treated with the PBE0 hybrid functional in combination with the Def2-TZVP basis set.\cite{10.1063/1.478522,B508541A} Geometry optimizations were carried out via a self-consistent Quasi-Newton approach employing the BFGS algorithm. A TightSCF convergence criterion was applied, corresponding to a total energy change threshold of 1.0e$^{-08}$ Eh and a one-electron integral threshold of 2.5e$^{-11}$ Eh. Van der Waals interactions were accounted for using Grimme’s DFT-D3(BJ) empirical dispersion correction. The ELF and LOL functions were calculated and visualized using the Multiwfn program.\cite{https://doi.org/10.1002/jcc.22885}

\section{Results}

The most stable structures of B$_8$Cu$_3^-$ clusters are identified using a modified basin-hopping (MBH) structure search method, following approaches described in previous works.\cite{D0CP06179D,D4CP04444D} Eight initial structures are randomly perturbed multiple times and subjected to the Metropolis criterion until hundred structures are selected. These structures are then evaluated through single-point energy calculations and ranked accordingly. The lowest-energy configurations are fully optimized and used to seed the next generation. The MBH method enhances standard random perturbations by incorporating random exchanges of atomic species, enabling a more efficient exploration of the potential energy surface of binary alloys.

\begin{figure}[h!]
  \begin{tabular}{ccc}
      \includegraphics[scale=0.3]{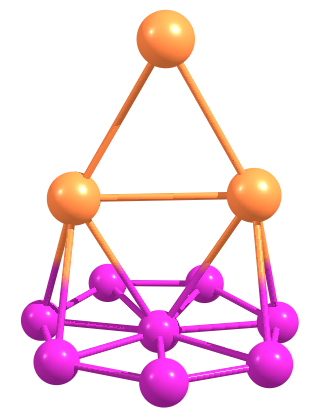} &
      \includegraphics[scale=0.24]{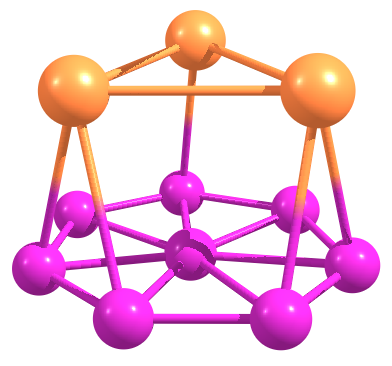} &
      \includegraphics[scale=0.3]{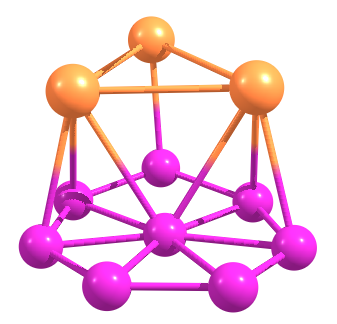} \\
      0.0, $C_s$ & 3.0, $C_s$ & 34.1, $C_s$ \\
      \end{tabular}
       \caption{\label{figure_struc}Lowest energy structures for B$_8$Cu$_{3}^-$ clusters. For each structure, the relative energy (in kcal/mol) is given. }
\end{figure}

The results reveal that the lowest-energy structure of the B$_8$Cu$_3^-$ cluster corresponds to a vertical Cu$_3$ triangle supported on a B$_8$ wheel-like geometry ({\bf 8M3.1}) in the singlet state. Similar classes of boron wheel structures have been previously reported.\cite{C2CP42218B,doi:10.1021/acsomega.6b00159,D4CP04444D} The next lowest-energy isomer ({\bf 8M3.2}) adopts a comparable arrangement but with a horizontal Cu$_3$ unit atop the B$_8$ ring, and lies 3.0 kcal/mol higher in energy. Another isomer, {\bf 8M3.3}, is also similar but has the Cu$_3$ motif displaced from the central site (Figure~\ref{figure_struc}). {Moreover, the results are further confirmed using the {$\omega$}B97x-D3/Def2-TZVP level, which showed the same trend but in a more pronounced manner, helping to validate our findings} (see Table~\ref{table_R1}).

\begin{table}[h!]
\caption{\label{table_R1}{Relative energies (in kcal/mol) of the lowest energy structures of B$_8$Cu$_3^-$ clusters computed at different DFT levels. For each functional, the most stable configuration is specified (*). The clusters are labeled by the {\bf nM3.y} notation, where {\bf n} is the number of boron atoms, {\bf M3} denotes three Cu atoms, and {\bf y} stands for the isomer number.\cite{b7cr2clusters,b7al2,al12m,pt5v} }}
\small
\centering
\begin{tabular}{ c c c c }
 {Label}  & m & \small{PBE0/Def2-TZVP} & \small{{$\omega$}B97X-D3/Def2-TZVP} \\
\hline
 \textbf{8M3.1} & 1 & 0.0*  & 0.0*     \\
\textbf{8M3.2} & 1 & 3.0 & 5.6    \\
\textbf{8M3.3} & 1 & 34.1  & 16.0      \\
\hline
\end{tabular}
\end{table}

\begin{figure}[ht]

 \resizebox*{0.40\textwidth}{!}{\includegraphics{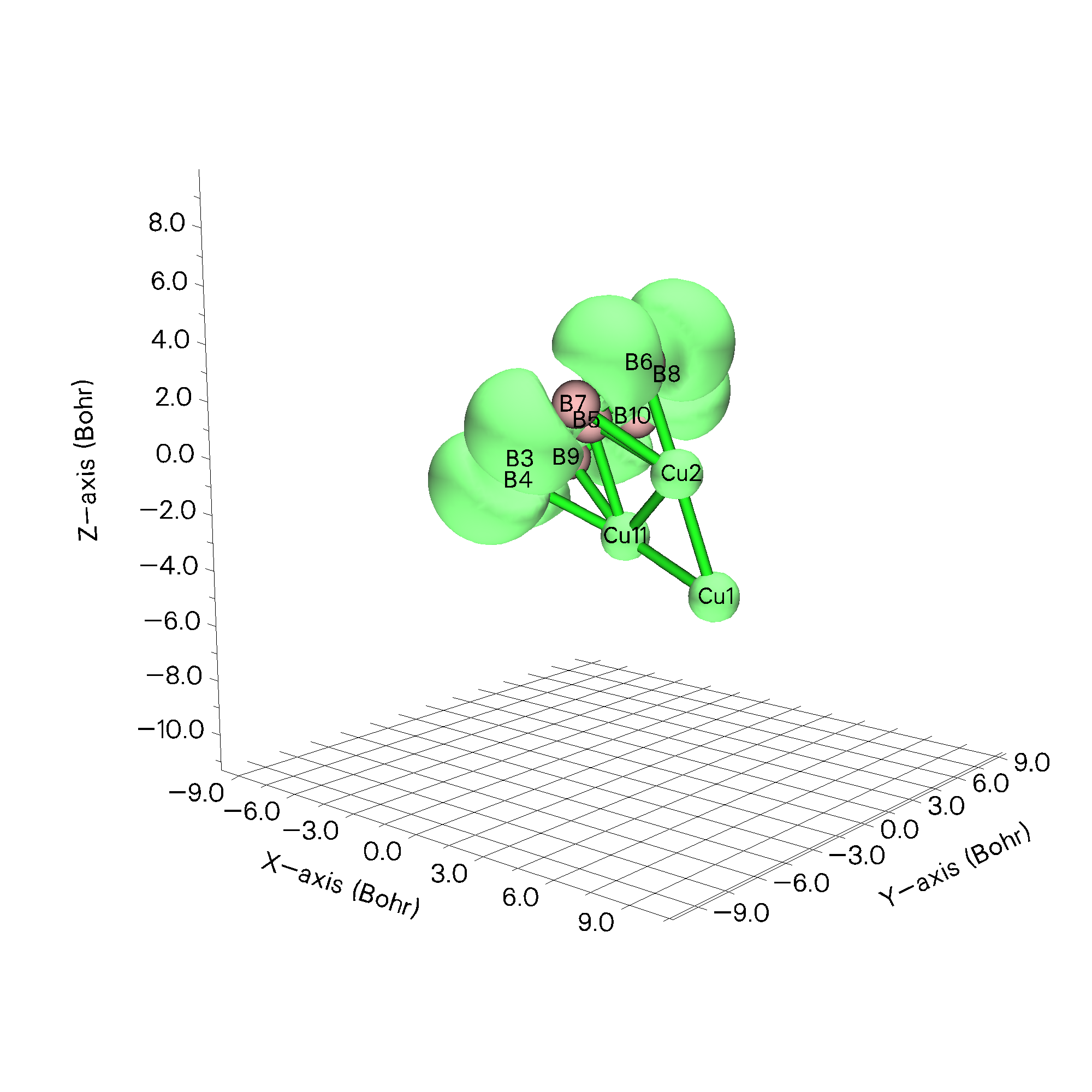}} 
\resizebox*{0.40\textwidth}{!}{\includegraphics{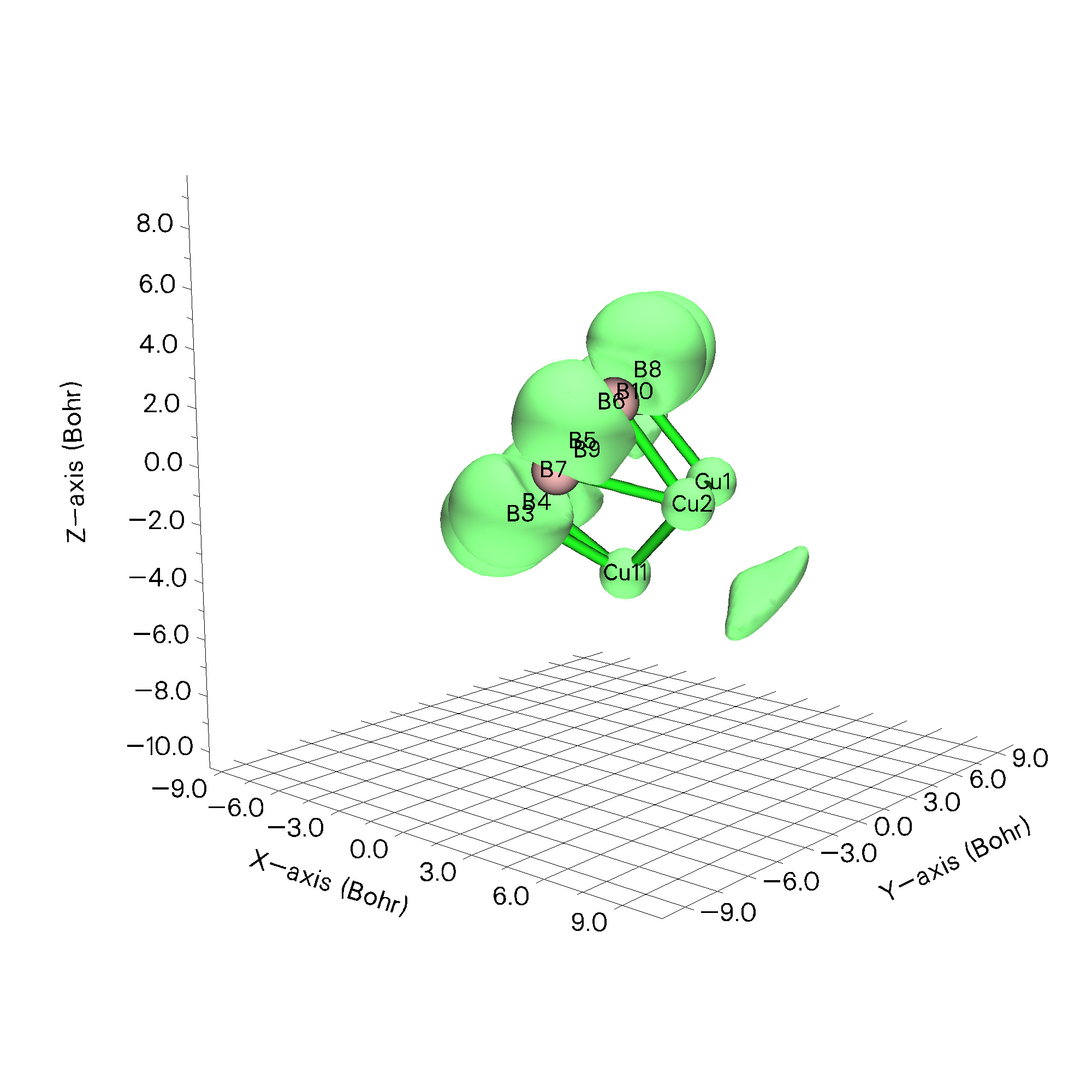}} 
    \caption{\label{figure_2}
Electron localization function (ELF) isosurfaces (0.70 a.u.) for two isomers of the B$_8$Cu$_3^-$ cluster. (Top) The most stable isomer ({\bf 8M3.1}) shows a vertical Cu$_3$ triangle strongly interacting with the B$_8$ wheel. (Bottom) A higher-energy isomer ({\bf 8M3.2}) with a horizontal Cu$_3$ arrangement exhibits weaker Cu–B interactions and less localized electron density between Cu and B atoms.}
\end{figure}

\begin{table}[ht!]
{
\caption{\label{table_1}{Mulliken population analysis of the most stable B$_8$Cu$_3^-$ clusters. }}
\small
\def\arraystretch{1.1}
\begin{tabular}{p{1.2cm}p{1.5cm}p{1.5cm}p{0.5cm}p{1.5cm}p{1.5cm}}
\hline
 &  \multicolumn{2}{c}{\bf 8M3.1} &  &  \multicolumn{2}{c}{\bf 8M3.2} \\ \cline{2-3} \cline{5-6} 
   Atom   &   Pop. &  Charge &    &  Pop. &  Charge \\
   1(Cu)  &   29.305 &  -0.305 &  & 29.196   & -0.196\\
   2(Cu)  &   29.148 &  -0.148 & & 29.202   & -0.202\\
   3(B )  &   5.131  &  -0.131 & & 5.024    & -0.024\\
   4(B )  &   5.218  &  -0.218 & & 4.930    &  0.069\\
   5(B )  &   4.529  &   0.470 & & 5.235    & -0.235\\
  6(B )   &   4.939  &   0.060 & & 4.910    &  0.089\\
  7(B )   &   5.218  &  -0.218 & &  5.090   & -0.090\\
  8(B )   &   5.221  &  -0.221 & & 5.141    & -0.141\\
  9(B )   &   4.903  &    0.096 & & 5.145   & -0.145\\
 10(B )   &   5.238  &   -0.238 & & 4.920   &  0.079\\
 11(Cu)   &  29.144  &   -0.144 & & 29.204  & -0.204\\
\hline
\end{tabular}
	}
\end{table}

\begin{figure*}[htb!]
{
\small
\def\arraystretch{1.1}
\begin{tabular}{cc}
\resizebox*{0.30\textwidth}{!}{\includegraphics{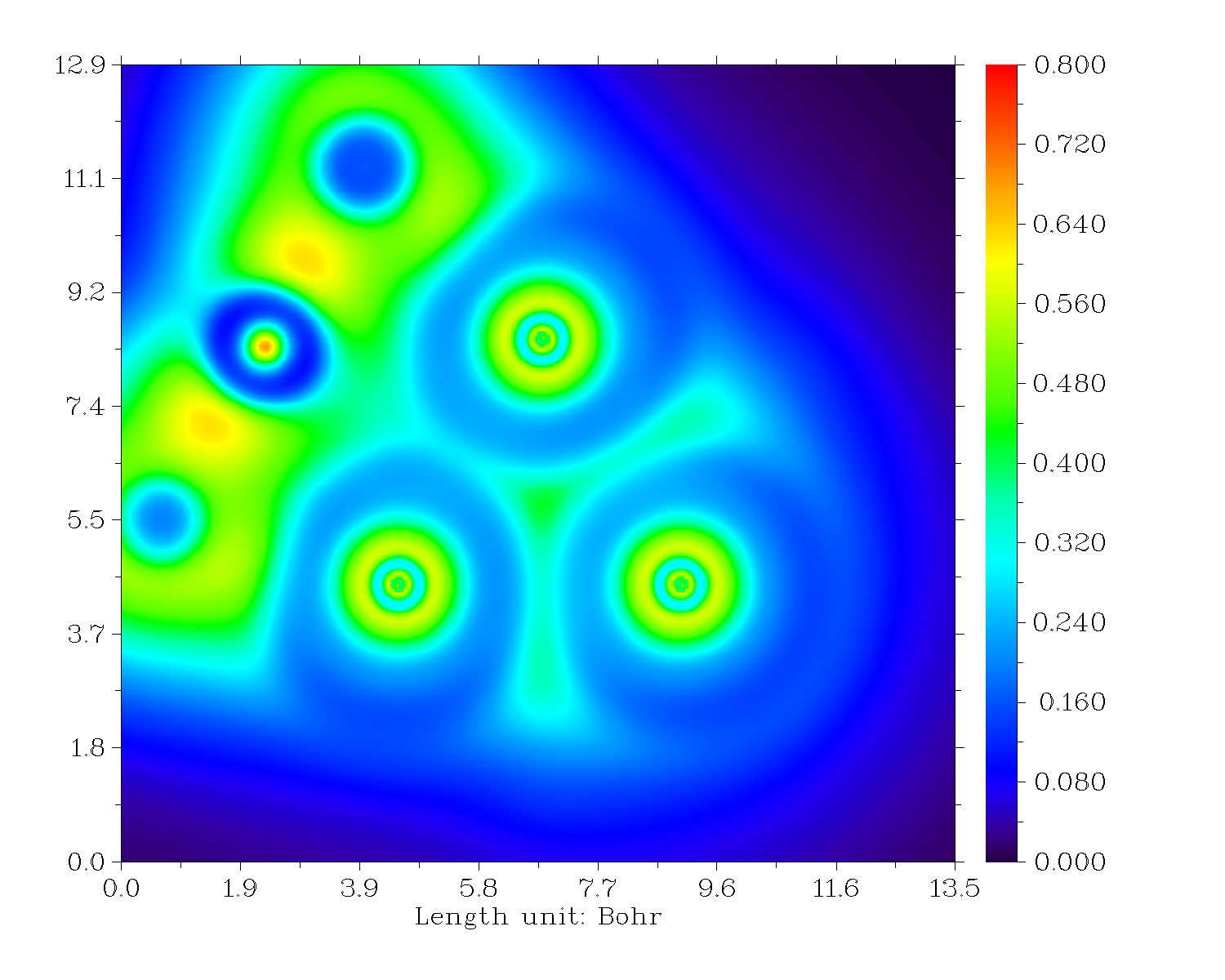}} &
\resizebox*{0.30\textwidth}{!}{\includegraphics{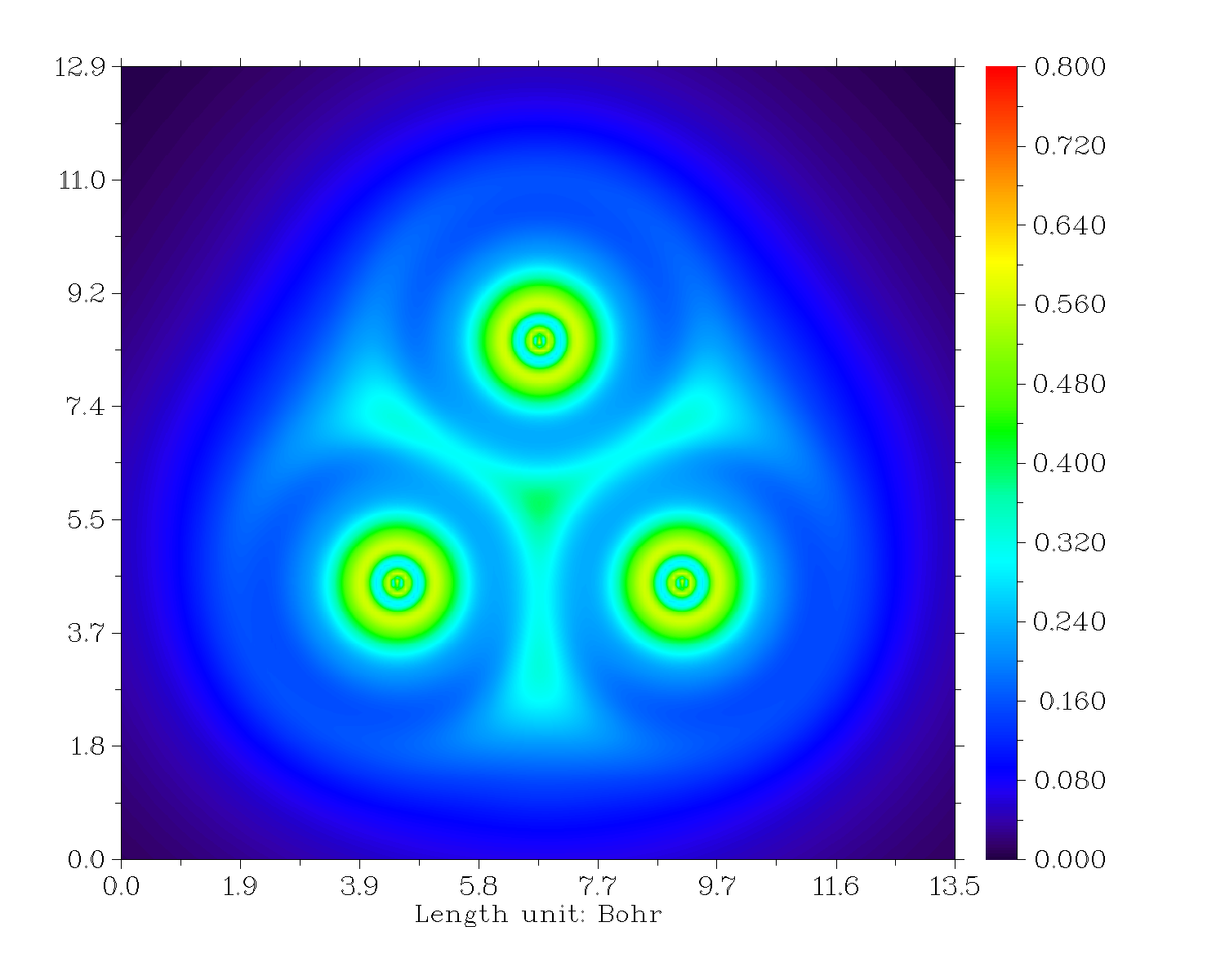}} \\
\end{tabular}
	}
    \caption{\label{fig_3}{Localized Orbital Locator (LOL) maps for the {\bf 8M3.1} and {\bf 8M3.2} isomers calculated at the PBE0/Def2TZVP level. The color scale indicates the degree of electron localization, with red denoting high localization and blue indicating low localization.}}
\end{figure*}

To gain insight into the bonding stability of the clusters, we evaluated the electron localization function (ELF) analysis of the clusters, which, provides valuable insights into the electronic structure and stability of the isomers,\cite{C7CP01740E,MAI201987,2025stabreactdoubleicosahedron} as depicted in Figure~\ref{figure_2}. The electron localization function (ELF) plots for the two B$_8$Cu$_3^-$ isomers provide insight into the nature of bonding and electron delocalization in these clusters. In the most stable isomer (first image), the ELF indicates a significant localization between the Cu atoms and the B$_8$ wheel, particularly along the vertical Cu$_3$ triangle, suggesting strong Cu–B interactions and possibly multicenter bonding that stabilizes the structure. The electron density appears to be more evenly distributed around the B atoms, consistent with a delocalized aromatic-type bonding within the boron wheel. In contrast, the second isomer (second image), which features a horizontal Cu$_3$ motif, shows less pronounced ELF features between Cu atoms and the B$_8$ core, and the localization appears more fragmented. This suggests weaker Cu–B interactions and possibly reduced electronic delocalization compared to the first isomer, which correlates with its higher energy and lower stability. Overall, the ELF analysis reinforces the structural preference for a vertically aligned Cu$_3$ motif interacting effectively with the delocalized B$_8$ framework.

The Mulliken population analysis presented in Table~\ref{table_1} provides insights into the electronic structure and charge distribution of the two most stable B$_8$Cu$_3^-$ isomers. For the lowest-energy isomer, {\bf 8M3.1}, the Cu atoms exhibit more negative Mulliken charges (e.g., –0.305 for Cu1) compared to those in {\bf 8M3.2}, indicating a greater degree of electron density localization on the copper centers. In contrast, the B atoms in {\bf 8M3.1} show a more polarized distribution, with some atoms (e.g., B5) acquiring significant positive charge (+0.470), suggesting localized electron depletion likely due to bonding interactions with the Cu triangle. This pronounced charge separation may contribute to the enhanced thermodynamic stability of {\bf 8M3.1}. Meanwhile, the more uniform and less extreme charge distribution in {\bf 8M3.2} implies weaker or more delocalized interactions, consistent with its slightly higher energy (3.0 kcal/mol above {\bf 8M3.1}). Overall, the charge analysis supports the conclusion that stronger and more directed Cu–B interactions in {\bf 8M3.1} play a key role in stabilizing this isomer.

Moreover, the localized orbital locator (LOL) function calculated via density functional theory (DFT) for the most stable B$_8$Cu$_3^-$ cluster isomers, are shown in shown in Figure~\ref{fig_3}. The LOL function provides a spatial visualization of electron localization, which is valuable for understanding the bonding characteristics within the cluster. It reveals key insights into the bonding nature of two B$_8$Cu$_3^-$ cluster isomers, with a focus on the Cu$_3$ trimer region. In the most stable isomer, the LOL map displays pronounced regions of electron localization around the Cu atoms and between Cu and adjacent B atoms, indicating strong, well-defined bonding interactions. In contrast, the less stable isomer shows a more diffuse distribution of electron density and weaker localization within the Cu$_3$ unit, suggesting diminished bonding strength. These differences in electron localization directly correlate with the relative stabilities of the two structures and highlight the role of Cu$_3$-centered bonding in stabilizing the cluster. These results provide a comprehensive understanding of the bonding and stability mechanisms in B$_8$Cu$_3^-$ clusters, offering valuable insights for the design of novel boron-based materials and transition metal-doped nanostructures.\cite{RODRIGUEZKESSLER2025122376}

\section{Conclusions}
In summary, our DFT investigation of B$_8$Cu$_3^-$ clusters reveals that the most stable isomer adopts a vertical Cu$_3$ triangle configuration supported by a B$_8$ wheel, significantly lower in energy than the horizontal counterpart. The combined analyses of the electron localization function (ELF), Mulliken population, and localized orbital locator (LOL) confirm strong Cu–B interactions and enhanced electron delocalization in the most stable structure. These features point to the presence of possible multicenter bonding, particularly involving the Cu$_3$ unit, as a key stabilizing factor. This study provides fundamental insights into the electronic structure and bonding mechanisms of boron–copper nanoclusters, offering guidance for the rational design of stable, transition-metal-doped boron-based materials.


\section{Acknowledgments}
P.L.R.-K. would like to thank the support of CIMAT Supercomputing Laboratories of Guanajuato and Puerto Interior. 



\bibliographystyle{unsrt}
\bibliography{mendelei.bib}
\end{document}